\pdfoutput=1

\documentclass[11pt,twoside,a4paper,cmspaper,final,collab]{cms-tdr}
\def\svnVersion{389300}\def\cmsCernNoTag{CERN-EP/2016-190}\def\cmsCernDate{\today}\def\cmsMessage{Published in the European Physical Journal C as \href{http://dx.doi.org/10.1140/epjc/s10052-016-4573-z}{\doi{10.1140/epjc/s10052-016-4573-z.}}}

\begin{document}\cmsNoteHeader{SMP-14-020}

\hyphenation{had-ron-i-za-tion}
\hyphenation{cal-or-i-me-ter}
\hyphenation{de-vices}
\RCS$Revision: 374439 $
\RCS$HeadURL: svn+ssh://svn.cern.ch/reps/tdr2/papers/SMP-14-020/trunk/SMP-14-020.tex $
\RCS$Id: SMP-14-020.tex 374439 2016-11-21 13:27:37Z tmperry $
\newcolumntype{x}{D{,}{\pm}{-1}}
\newlength\cmsFigWidth
\ifthenelse{\boolean{cms@external}}{\setlength\cmsFigWidth{0.49\textwidth}}{\setlength\cmsFigWidth{0.75\textwidth}}
\ifthenelse{\boolean{cms@external}}{\providecommand{\cmsLeft}{top}}{\providecommand{\cmsLeft}{left}}
\ifthenelse{\boolean{cms@external}}{\providecommand{\cmsRight}{bottom}}{\providecommand{\cmsRight}{right}}
\newcommand{\ZZ}{\cPZ{}\cPZ\xspace}
\newcommand{\WW}{\PWp{}\PWm\xspace}
\newcommand{\WZ}{\PW{}\cPZ\xspace}
\newcommand{\ttjets}{\ensuremath{\ttbar}+\text{jets}\xspace}
\newcommand{\Wlnbb}{\ensuremath{\PW(\ell\nu)}+\bbbar\xspace}
\newcommand{\Wjets}{\ensuremath{\PW}+\text{jets}\xspace}
\newcommand{\Vjets}{\ensuremath{\mathrm{V}}+\text{jets}\xspace}
\newcommand{\Gjets}{\ensuremath{\gamma}+\text{jets}\xspace}
\newcommand{\MT}{\ensuremath{M_\mathrm{T}}\xspace}
\newcommand{\Wudscg}{\PW{}+\PQu{}\PQd{}\PQs{}\PQc{}\Pg{}\xspace}
\newcommand{\Wcc}{\PW{}+\ccbar{}\xspace}
\newcommand{\Wbb}{\PW{}+\bbbar{}\xspace}

\cmsNoteHeader{SMP-14-020}
\title{Measurement of the production cross section of a W~boson in association with two b jets in pp collisions at $\sqrt{s} = 8\TeV$}

\date{\today}
\titlerunning{Measurement of the production cross section of a W boson with two b jets}
\abstract{
The production cross section of a W~boson
 in association
 with two b jets is measured using a
 sample of proton-proton collisions at
 $\sqrt{s} = 8\TeV$ collected by the CMS experiment at the CERN LHC.
The data sample corresponds to an integrated
 luminosity of 19.8\fbinv.
The W~bosons are reconstructed via their
 leptonic decays, {$\PW\to\ell\nu$},
 where  {$\ell=\mu$ or \Pe}.
The fiducial region studied contains exactly one
 lepton with transverse momentum {$\pt^{\ell}>30\GeV$}
 and pseudorapidity {$\abs{\eta^{\ell}}<2.1$},
 with exactly two b jets with {$\pt>25\GeV$}
 and {$\abs{\eta}<2.4$}
 and no other jets with {$\pt>25\GeV$}
 and {$\abs{\eta}<4.7$}.
The cross section is measured to be
 {$\sigma ( {{\Pp\Pp}} \to {\PW} (\ell\nu)$+$\bbbar)= 0.64 \pm 0.03\stat \pm 0.10\syst \pm 0.06\thy \pm 0.02\lum\unit{pb}$},
 in agreement with standard model predictions.
}
\hypersetup{%
pdfauthor={CMS Collaboration},%
pdftitle={Measurement of the production cross section of the W boson in association with two b jets in pp collisions at sqrt(s) = 8 TeV},%
pdfsubject={SMP},%
pdfkeywords={CMS, physics, SMP}
}
\maketitle

\section{Introduction}

The measurement of $\PW$ or $\PZ$~boson
 production in association with b quarks in proton-proton
 collisions provides important input for refinement of
 calculations in perturbative quantum chromodynamics
 and is also relevant for searches and measurements.
In particular, these processes constitute
 a background to the experimental measurement of a
 standard model (SM)
 Higgs boson in which the Higgs boson decays into a $\bbbar$ pair
 in association with a vector boson.
The discovery by the ATLAS and CMS
 Collaborations at the CERN LHC of a neutral boson with a mass of about
 $125\GeV$~\cite{Aad:2015zhl,Khachatryan:2014jba,Chatrchyan:2012ufa,Aad:2012tfa}
 motivates further studies to
 establish the nature of the boson and determine its
 coupling to bottom quarks.
Furthermore, different models based on extensions of the
 Higgs sector are being compared with LHC data
 using final states composed of leptons and b jets.
In this context, a better understanding of the b hadron
 production mechanism and the kinematic properties of associated jets
 is required to refine
 the background predictions and increase the sensitivity to new physics.
Throughout this paper,
 hadronic showers originating from
 bottom or anti-bottom quarks
 are referred to as b jets, and b-tagged jets
 are the reconstructed objects either in simulation or data
 that have been identified as such.

The production of $\PW$~\cite{Chatrchyan:2013uza,Aad:2013vka}
 or $\PZ$~\cite{Chatrchyan:2012vr,Chatrchyan:2014dha,Chatrchyan:2013zja,Aad:2011jn,Aad:2014dvb}~bosons
 in association with b jets has been
 measured at the LHC using pp collisions at
 $\sqrt{s}=7\TeV$ using data samples corresponding to
 an integrated luminosity of up to 5\fbinv,
 and at the Fermilab Tevatron~\cite{WbbTevD0,WbbTevCDF} using
 proton-antiproton collisions at $\sqrt{s}=1.96\TeV$.
This analysis extends previous measurements of the
 \Wbb production cross section~\cite{Chatrchyan:2013uza}
 and uses data at $\sqrt{s}=8\TeV$ collected with the CMS detector,
 corresponding to an integrated luminosity of 19.8\fbinv~\cite{ref:CMSLumiCalc}.
Whereas the previous CMS analysis used only the muon decay channel,
 this analysis uses both muon and electron decay modes.

\section{CMS detector}
The central feature of the CMS apparatus is a superconducting solenoid of 6\unit{m} internal diameter,
 providing a magnetic field of 3.8\unit{T}.
Within the solenoid volume are a silicon pixel and strip tracker,
 a lead tungstate crystal electromagnetic calorimeter (ECAL),
 and a brass and scintillator hadron calorimeter (HCAL),
 each composed of a barrel and two endcap sections.
Muons are measured in gas-ionization detectors embedded
 in the steel flux-return yoke outside the solenoid.
Extensive forward calorimetry complements the
 coverage provided by the barrel and endcap detectors.
A more detailed description of the CMS detector,
 together with a definition of the coordinate system used
 and the relevant kinematic variables, can be found in Ref.~\cite{Chatrchyan:2008zzk}.

\section{Event selection and reconstruction}
\label{sec:event_selection_and_reconstruction}

 The {$\PW\to \Pgm\nu_\Pgm$}
 ({$\PW\to\Pe\nu_\Pe$}) events are selected using
 single-muon (single-electron) triggers that require a
 loosely isolated muon (electron)
 with transverse momentum {$\pt>24\,(27)\GeV$}
 and pseudorapidity {$\abs{\eta}<2.1\,(2.5)$}.

Individual particles emerging from each collision are reconstructed with the
 particle-flow (PF) technique~\cite{CMS-PAS-PFT-10-001,CMS:2009nxa}.
This approach uses the information from all subdetectors to identify and
 reconstruct individual particle candidates in the event, classifying
 them into mutually exclusive categories: charged and neutral hadrons, photons, electrons, and muons.

Muons are reconstructed by combining the information
 from the tracker and the muon spectrometer~\cite{ref:muon_detector,CMS-PAS-MUO-10-002}.
Electrons are reconstructed by combining the information
 from the tracker and the calorimeter~\cite{ref:ele_detector}.
Both the muon and the electron candidates are required to have
 {$\pt>30\GeV$} and {$\abs{\eta}<2.1$} to ensure
 that the triggers are fully efficient.
They are also required to originate
 from the primary vertex of the event,
 chosen as the vertex with the highest $\sum \pt^2$
 of the charged particles associated with it.
Furthermore, the leptons must be isolated, where
 the isolation variable is defined as
\ifthenelse{\boolean{cms@external}}
{
\begin{multline}
\label{eq:iso}
I =\frac{1}{\pt^\ell}\biggl[ \sum\pt^{\text{charged}}\\+{\max}\Bigl(0,\textstyle{\sum}\pt^{\gamma}+\textstyle{\sum}\et^{\text{neutral}}-0.5\,\textstyle{\sum} \pt^{\mathrm{PU}}\Bigr)\biggr],
\end{multline}
}{
\begin{equation}
\label{eq:iso}
I =\frac{1}{\pt^\ell}\left[ \sum\pt^{\text{charged}}+{\max}\Bigl(0,\textstyle{\sum}\pt^{\gamma}+\textstyle{\sum}\et^{\text{neutral}}-0.5\,\textstyle{\sum} \pt^{\mathrm{PU}}\Bigr)\right],
\end{equation}
}
with the sums running over PF candidates
 in a cone of size $\Delta R < 0.4\,(0.3)$ around the muon (electron) direction,
 where {$\Delta R = \sqrt {\smash[b]{ (\Delta \eta)^2 + (\Delta \phi)^2}}$},
 and $\phi$ is the azimuthal angle in radians.
The first three sums are over
 charged hadron candidates associated with the primary vertex,
 photon candidates, and neutral hadron candidates respectively.
The definition of the isolation includes a correction for
 additional pp interactions, referred to as pileup,
 which is proportional to the scalar~\pt sum of charged particles
 not associated with the primary vertex in the isolation cone
({$\sum \pt^{\mathrm{PU}}$}).
The selected muons (electrons) are required to
 have {$I < 0.12\,(0.10)$}.

Missing transverse momentum in the event, $\ptvecmiss$,
 is defined as the negative vector sum of
 the $\vec{\pt}$ of all PF candidates
 in the event.
It is combined with the $\vec{\pt}$ of a muon or electron
 passing the identification and isolation requirements
 to compute the transverse mass, \MT, of the \PW~boson candidate.
The \MT variable is a
 natural discriminator against non-$\PW$ final states,
 such as quantum chromodynamics (QCD) multijet events,
 that have a lepton candidate and $\ptvecmiss$,
 but a relatively low value of $\MT$.
The result for $\ptvecmiss$ is corrected for noise in the
 ECAL and HCAL using the method described in Ref.~\cite{WZCMS:2010}.
Corrections to minimize the effect of the
 pileup are also included~\cite{CMS:8TeVMET}.

Jets are constructed using the anti-\kt
 clustering algorithm~\cite{Cacciari:2008gp}
 with a radius parameter of 0.5,
 as implemented in the \FASTJET package~\cite{fastjet1,fastjet2}.
Jet clustering is performed using individual
 particle candidates reconstructed with the PF technique.
Jets are required to pass identification
 criteria that eliminate jets originating from
 noisy channels in the HCAL~\cite{Chatrchyan:2009hy}.
Jets from pileup interactions are rejected by
 requiring that the jets originate at
 the primary interaction vertex.
Small corrections to the relative and absolute jet
 energy calibrations of the detector are
 applied as a function of the $\pt$ and $\eta$  of the jet~\cite{cmsJEC}.

The combined secondary vertex (CSV)
 b tagging algorithm~\cite{CMS-PAS-BTV-13-001,cmsBTAGPAPER} exploits
 the long lifetime and relatively large mass of b hadrons to provide
 b jet identification.
The CSV algorithm combines information about impact parameter
 significance, secondary vertex kinematic properties, and jet kinematic
 properties in a likelihood-ratio discriminator.
The identification of b jets (b tagging)
 is made by imposing a minimum threshold on
 the CSV discriminator value.
In this analysis, b-tagged jets are required
 to pass a threshold with an efficiency
 of 40\% in the signal phase space and
 a misidentification probability of
 0.1\% (1\%) for light (charm) jets.
Jets are corrected for the difference in efficiency
 between data and simulation using scale factors
 dependent on the \pt of the jet.

\section{Simulated samples}

After all selection requirements
 detailed in Section \ref{sec:analysis_strategy}  are applied,
 the contributing background processes to the overall yield are the
 associated production of a massive vector boson and jets
 (\Vjets where V = \PW\ or \PZ),
 as well as production of diboson ($\WW$, $\WZ$, $\ZZ$), $\ttbar$,
 single top quark, \Gjets, and QCD multijet events.
These background contributions are estimated from simulation,
 except for the QCD background, which
 is estimated from data as described
 in Section \ref{sec:analysis_strategy}.

Simulated samples of \Vjets, \Gjets
 and  \ttjets are generated at tree-level with
 {\MADGRAPH~5.1} \cite{Alwall:2011uj,ref:MadEvent}
 using the CTEQ6L~\cite{ref:CTEQ66} parton distribution function (PDF) set.
These samples are interfaced with
{\PYTHIA~6.4}~\cite{Sjostrand:2006za} for hadronization
  using the Z2* tune
 for the underlying event.
The most recent {\PYTHIA} Z2* tune
 is derived from the Z1 tune~\cite{Field:2010bc},
 which uses the CTEQ5L PDF set,
 whereas Z2* adopts CTEQ6L~\cite{ref:CTEQ66}.
The \kt-MLM~\cite{Alwall:2007fs,Alwall:2008qv}
 matching scheme is used.
For the signal distributions,
 the shapes are taken from a
 dedicated high-statistics generated sample
 of exclusive \Wbb. 
The normalization is obtained from the \Wbb component
 of an inclusive \Wjets sample by
 separating the \Wjets simulated
 sample into three subsamples labeled as
 \Wbb, \Wcc, and \Wudscg,
 which are defined below.
If an event contains a bottom jet
 from the matrix element or parton shower,
 it is categorized as \Wbb.
A bottom quark at generator level requires
 the presence of a bottom hadron within a cone of
 radius $\Delta R=0.4$ with respect to the jet axis.
The jets are constructed using generator-level information using
 all stable particles in the event, excluding neutrinos.
Jets with a distance smaller than $\Delta R = 0.5$ with respect to a
 lepton are removed from the event.
If an event does not contain any b jet, but
 an even, nonzero number of charm jets,
 again from the matrix element or
 parton shower, it is categorized as \Wcc.
The remaining events are categorized as \Wudscg.
The energy of the selected leptons at the generator level is corrected
 for final-state radiation by summing
 the four-momenta of
 all the photons generated within a cone of radius $\Delta R = 0.1$
 around the lepton.
Leptons that do not originate from the primary vertex
 are not considered for selection.

Single top quark event samples are generated
 at next-to-leading order (NLO) with
 {\POWHEG~2.0} \cite{Nason:2004rx,Frixione:2007vw,ref:POWHEG_T_SandTchannel,ref:POWHEG_T_tWchannel}
 using the CTEQ6M PDF set.
Hadronization is performed using
 {\PYTHIA~6.4} with the Z2* tune.
Diboson samples are generated and hadronized
 with {\PYTHIA~6.4} at leading order (LO)
 using the CTEQ6L PDF set and the Z2* tune.

The cross sections for the \Vjets processes
 are normalized using the predictions
 for inclusive \PW\ and \PZ~boson production from
 \FEWZ~3.1~\cite{Gavin:2012sy,ref:FEWZ3.1}
 evaluated at next-to-next-to-leading order (NNLO).
The cross section for \Gjets is evaluated at LO
 using \MADGRAPH
 with the CTEQ6L PDF set.
Single top quark and diboson production cross sections are
 normalized to the NLO
 cross section predictions from $\MCFM~7.0$~\cite{Campbell:2010ff, Badger:2010mg}
 using the MSTW2008 NLO PDF set~\cite{Martin:2009iq}.
The $\ttbar$ cross section used is ${241.5\pm8.5\unit{pb}}$, and
 was determined from data collected by the ATLAS and CMS
 experiments~\cite{CMS:2014gta,Aad:2014kva,Chatrchyan:2013faa}
 at the LHC at $\sqrt{s}=8\TeV$.

For all simulated processes, the detector response is simulated using a detailed
 description of the CMS detector based on  \GEANTfour~\cite{GEANT}.
The reconstruction of simulated events is performed with
 the same algorithms used for the data.

Events induced by additional simultaneous pp interactions
 are simulated using events generated with {\PYTHIA~6}.
During the 2012 data taking, the average pileup rate was
 21 interactions per bunch crossing;
 the simulated number of pileup interactions
 has been reweighted to match this distribution in the data.

\section{Analysis strategy\label{sec:analysis_strategy}}
The \Wbb yield is estimated using a binned maximum-likelihood fit
 to the $\MT$ distribution in the signal event sample.
With the exception of the multijet processes,
 the distributions and normalizations of all
 background contributions
 in the fit are taken from simulation.
Consequently, it is important
 to verify that
 the simulation describes the data.

The dominant background in the signal event sample arises
 from the $\ttbar$ process.
Therefore, the data and simulation
 are compared in two $\ttbar$-dominated control samples:
 one characterized by a pair of opposite flavor leptons
 ($\ttbar$-multilepton),
 and the other by the presence of three or more jets
 ($\ttbar$-multijet).
The simulation is reweighted to describe the data in the control regions
 and then is used to predict the $\MT$ distributions
 in the signal region.

The signal region contains a muon (electron)
 with $\pt > 30\GeV$,
 {$\abs{\eta} < 2.1$}, and satisfying {$I < 0.12\,(0.10)$}.
Exactly two b-tagged jets with
 {$\pt > 25\GeV$} and
 {$\abs{\eta} < 2.4$} are also required.
Events with additional leptons with
 {$\pt > 10\GeV$} and {$\abs{\eta} < 2.4$} or
 a third jet with {$\pt > 25\GeV$} and {$\abs{\eta} < 4.7$}
 are rejected.
The $\ttbar$-multijet sample is obtained
 using the same selection criteria as for the signal
 event sample,
 but requiring at least three jets in the event
 with {$\pt>25\GeV$} and {$\abs{\eta}<2.4$} instead of
 vetoing events that have more than two jets.
The $\ttbar$-multilepton sample uses
 similar selection criteria as the signal event sample;
 however, the lepton requirement is
 modified.
The event must contain two isolated leptons of
 different flavor,
 each with
 {$\pt>30\GeV$} and {$\abs{\eta}<2.1$}.
In the $\ttbar$-multilepton sample,
 the $\MT$ variable is calculated with respect to
 the electron in the electron channel
 and the muon in the muon channel.

The QCD background distributions in the \MT variable are
 estimated from data using event samples that
 pass all signal requirements,
 but requiring the muon (electron) is not isolated,
 {$I > 0.20\,(0.15)$}.
The resulting distributions are corrected for the presence of all other
 backgrounds, as estimated from simulation.
Their contribution is less than 1\% of the QCD background rate.
The QCD background normalization is adjusted in order to describe
 the number of data events at {$\MT<20\GeV$},
 after subtracting the non-QCD backgrounds
 obtained from simulation.

In the fiducial regions used in this analysis,
 no correlation is observed between $I$ and $\MT$
 in multijet events simulated with \PYTHIA6,
 so the use of an inverted isolation requirement to
 obtain the QCD background distribution is possible.
However, this is not the case for the $\Delta R$ distance between the
 two b-tagged jets, $\Delta R(\PQb,\PAQb)$, or the lepton~$\pt$.
The shape of the QCD distribution for these variables is therefore
 taken from an $\MT<30$\GeV sideband and validated against
 QCD multijet simulation.
The normalization of the QCD background
 in these variables is set to the final
 normalization resulting from the fit to the \MT variable,
 which was derived using the inverted isolation requirement.

The normalizations and distributions of the simulated backgrounds
 are allowed to vary in the fit
 within the uncertainties listed in Table~\ref{tab:input_unc_wbb}
 as described in Section \ref{sec:systematic_and_statistical_uncertainties}.
The uncorrelated normalization uncertainties
 are uncertainties in the cross section of
 the given sample.

Two major parameters in the simulations
 significantly affect the
 normalization of the simulated distributions:
 the b tagging efficiency and the jet energy scale (JES).
The control samples as well as the signal
 event samples show similar
 sensitivity to the b tagging efficiency,
 and its adjustment affects all the regions
 in a correlated manner.
Because $\ttbar$ production may have
 more than two jets in the final state,
 the rejection of events with a third jet makes it sensitive to JES.
The effect on the leading jets is moderate, but
 JES variations lead to significant migration of jets
 into and out of the veto region.
The $\ttbar$-multijet sample,
 since it has no veto on a third jet,
 is less sensitive to JES variations
 than the $\ttbar$-multilepton sample.
The variation in the JES changes the \Wbb yield
 in the signal region by less than 1\%.

The fit procedure consists of three consecutive steps
 in which the simulated distributions in two control samples
 and the event sample
 are fit to data using the $\MT$ variable,
 which is chosen because it has
 a well-known shape for  \Wjets production
 that allows for reliable signal extraction.
First, the fit is performed using the $\ttbar$-multijet sample.
It results in a correction of the
 b tagging efficiency,
 measured separately in the muon and electron
 channels and then combined.
The simulation is corrected using this result
 and the corrected simulated samples are
 fit to the data in the $\ttbar$-multilepton sample.
The result of the second step is used to adjust JES
 and as a result of this procedure, the simulation is
 expected to better describe the $\ttbar$ contribution.
The final step is to extract the number of
 \Wbb events from the fit to \MT in the signal event sample.

Similar results can be obtained by
 performing a simultaneous fit of the
 signal and the two control regions. 
We find that the b tagging efficiency
 correction and JES correction have
 opposite effects on the distributions
 and thus compensate for each other
 in a simultaneous fit,
 reducing its precision. 
Separating these effects in steps provides
 better understanding of underlying
 uncertainties and therefore more precise results.

\section{Systematic uncertainties}
\label{sec:systematic_and_statistical_uncertainties}

The main sources of the systematic uncertainties
 are listed in Table~\ref{tab:input_unc_wbb}.
The size of the variation is shown for each source,
 together with its effect on the measured cross section.
These are included in the fit.
Some of the uncertainties affect
 only the normalization in the respective contributions.
These include the uncertainties in the
 theoretical cross section for a given sample,
 which are uncorrelated between samples
 and are included as log-normal constraints on the rate.
The uncertainty due to the b tagging efficiency
 and the uncertainty due to the JES
 are observed to
 only affect the normalizations of the samples
 in the \MT variable.
The uncertainties that affect both the normalization and the
 shape of the $\MT$ distributions
 are listed in
 the table under ``Shape'' and are
 incorporated into the fit via binned distributions,
 which are obtained by varying the source
 of the given uncertainty and reprocessing the simulated sample.
Such uncertainties in the template are interpolated quadratically.

As a conservative estimate of the uncertainty in QCD multijet
 background, a 50\% uncertainty has been considered.
This results in an
 uncertainty of 2--3\% in the measured cross section.
The b tagging efficiency and JES rescaling uncertainties
 are taken from their respective fits.
The renormalization and factorization
 scales respectively are set at
 $\mu_{\mathrm{R}}=\mu_{\mathrm{F}}=m_\PW$,
 and the
 uncertainties on this choice are estimated
 from the change in
 acceptance found by varying
 $\mu_{\mathrm{R}}$ and $\mu_{\mathrm{F}}$
 up and down by a factor of two.
The PDF uncertainties are estimated from the change in
 acceptance found by varying the
 PDF set following the LHAPDF/PDF4LHC prescription~\cite{Buckley:2014ana,Botje:2011sn,Alekhin:2011sk,Ball:2012cx},
 considering PDF sets from the CTEQ, MSTW, NNPDF, and HERA Collaborations.

\begin{table}[htb]
\centering
\topcaption{
The main sources of systematic uncertainty
 in the \Wbb signal event sample.
The column labeled ``Variation'' indicates the bounds
 on the normalization change of a given sample
 due to a variation of the uncertainty by one
 standard deviation.
The last column indicates the contribution of
 the given systematic to the overall uncertainty
 in the measured cross section.
The uncertainty labeled ``b tag eff rescaling'' is
 the uncertainty associated with the rescaling
 of the b tagging efficiency.
UES refers to the scale of energy deposits
 not clustered into jets, and MES and EES
 refer to the muon and electron energy scales.
The uncertainty labeled as "Id/Iso/Trg" is the uncertainty
 associated with the efficiency of the lepton
 identification, isolation, and trigger.
The uncertainties in the integrated luminosity~\cite{ref:CMSLumiCalc} and
 in the acceptance due to PDF uncertainties
 and scale choices are
 not included in the fit, and are
 treated separately.
}
\label{tab:input_unc_wbb}
{\renewcommand{\arraystretch}{1.2}
\begin{tabular}{c|c|l|l|l}
\multicolumn{4}{c}{}                                                            & \multicolumn{1}{c}{Effect on the}\\[-5pt]
\multicolumn{4}{c}{}                                                            & \multicolumn{1}{c}{measured} \\[-5pt]
\multicolumn{2}{c}{}                                                         & \multicolumn{1}{c}{Uncertainty}   & \multicolumn{1}{c}{Variation} & \multicolumn{1}{c}{cross section} \\
\hline
\hline
 \multirow{11}{*}{\rotatebox{90}{Normalization}} & \multirow{8}{*}{\rotatebox{90}{Uncorrelated}} & \ttbar        & 3.5\%     & 3.8\%    \\
            {}                                   &           {}                                  & Single top    & 5.4\%     & 2.5\%    \\
            {}                                   &           {}                                  &    \Wudscg    & 13.2\%    & ${<}2$\%   \\
            {}                                   &           {}                                  &       \Wcc    & 13.2\%    & ${<}2$\%   \\
            {}                                   &           {}                                  &    Diboson    & 8.1\%     & ${<}2$\%   \\
            {}                                   &           {}                                  &  Drell--Yan   & 7.9\%     & ${<}2$\%   \\
            {}                                   &           {}                                  & $\gamma$+jets & 10.0\%    & ${<}2$\%   \\
            {}                                   &           {}                                  &        QCD    & 50\%      & 2--3\%  \\
\cline{2-5}
\noalign{\smallskip}
\cline{2-5}
            {}                                   &\multirow{6}{*}{\rotatebox{90}{Correlated}}    & b tag eff rescaling & 8.4\%   & 9.2\% \\
            {}                                   &           {}                                  & JES rescaling       & 0--6\%   & 3.8\% \\
\cline{1-1} \cline{3-5}
\multirow{4}{*}{\rotatebox{90}{Shape}}           &           {}                                  & UES           &  0--3\%   & ${<}2$\%  \\
            {}                                   &           {}                                  & MES           &  0--3\%   & ${<}2$\% \\
            {}                                   &           {}                                  & EES           &  0--3\%   & ${<}2$\% \\
            {}                                   &           {}                                  & Id/Iso/Trg    &  0--4\%   & ${<}2$\% \\
\hline
\hline
\multicolumn{2}{c}{}&            \multicolumn{1}{l|}{Luminosity}                                     & \multicolumn{2}{c}{2.6\%} \\
\multicolumn{2}{c}{}&            \multicolumn{1}{l|}{Scales ($\mu_{\mathrm{R}}$,$\mu_{\mathrm{F}}$)} & \multicolumn{2}{c}{10\%}   \\
\multicolumn{2}{c}{}&            \multicolumn{1}{l|}{PDF choice}                                     & \multicolumn{2}{c}{1\%}
\end{tabular}
}
\end{table}

\section{Results}
\label{sec:results}

\begin{figure*}[htbp]
\centering
\includegraphics[width=0.45\textwidth]{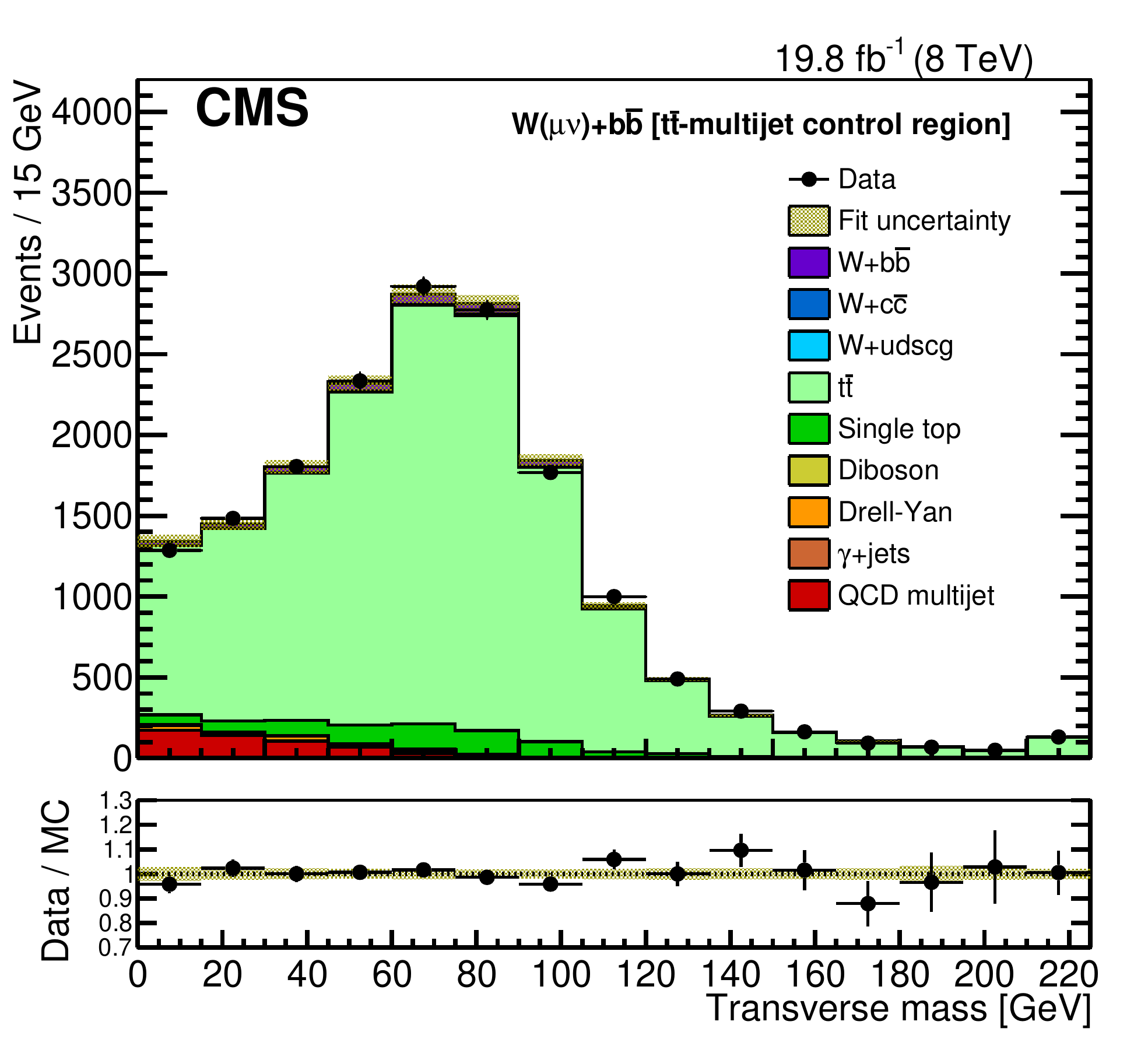}
\includegraphics[width=0.45\textwidth]{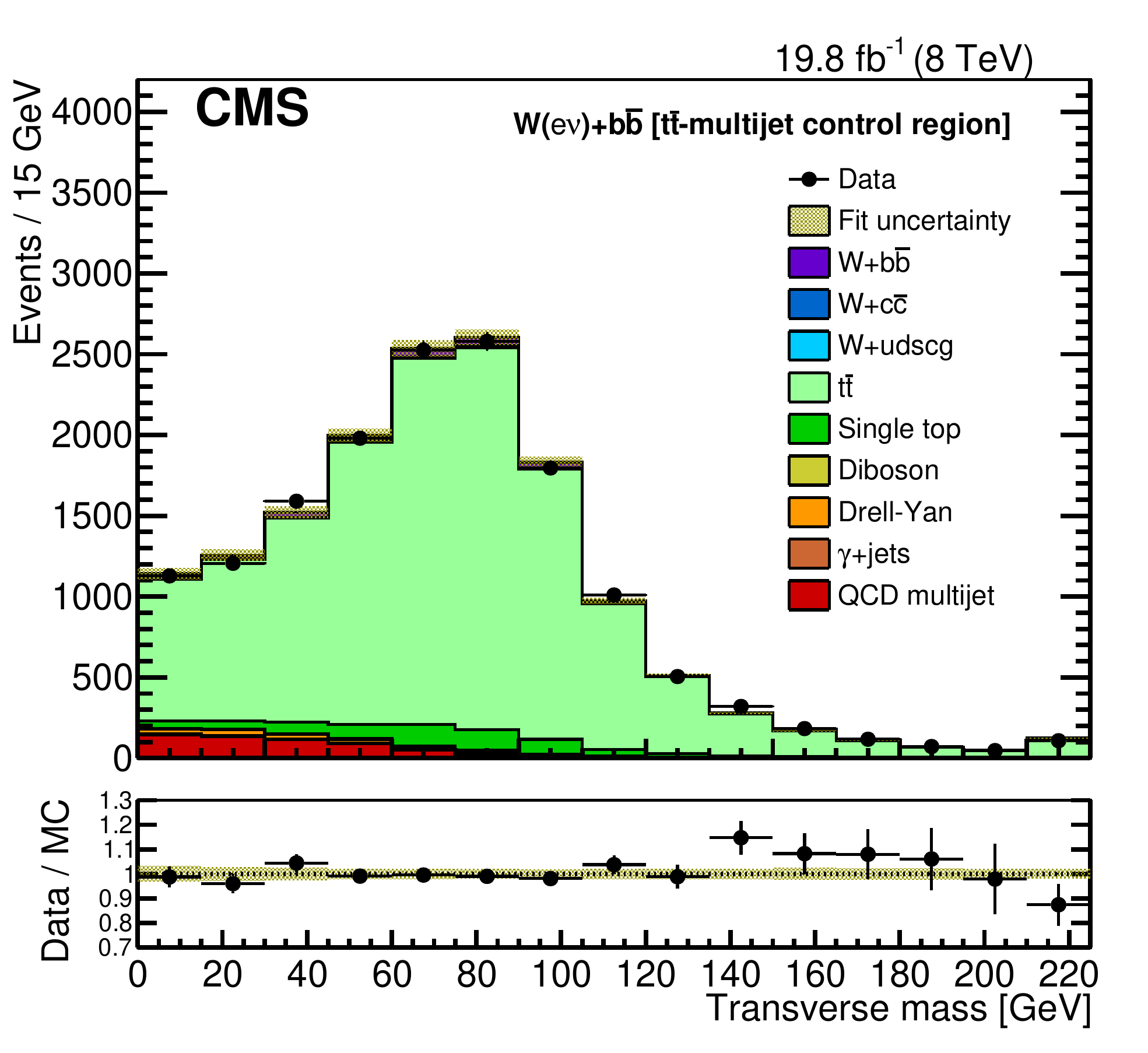}
\includegraphics[width=0.45\textwidth]{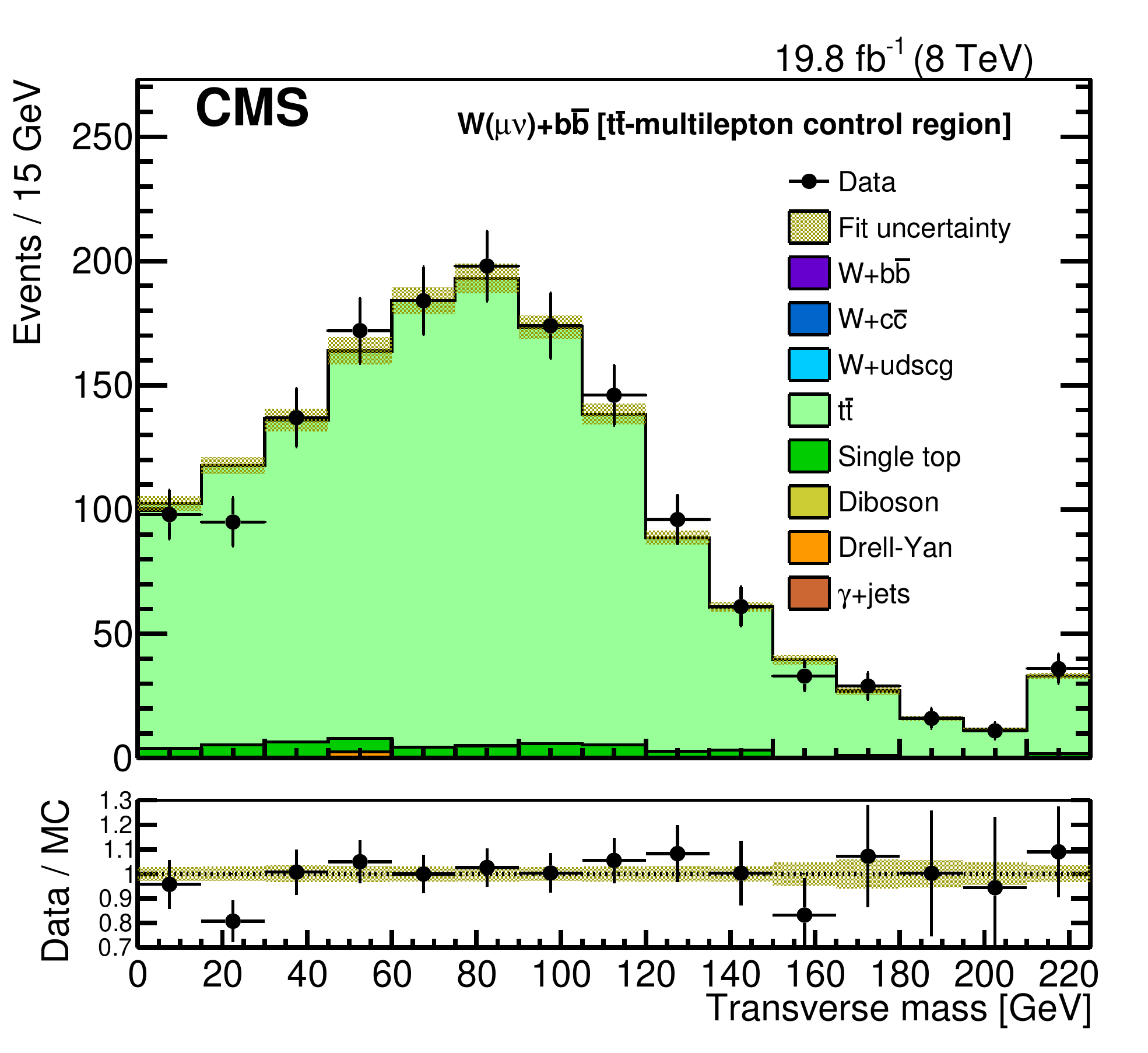}
\includegraphics[width=0.45\textwidth]{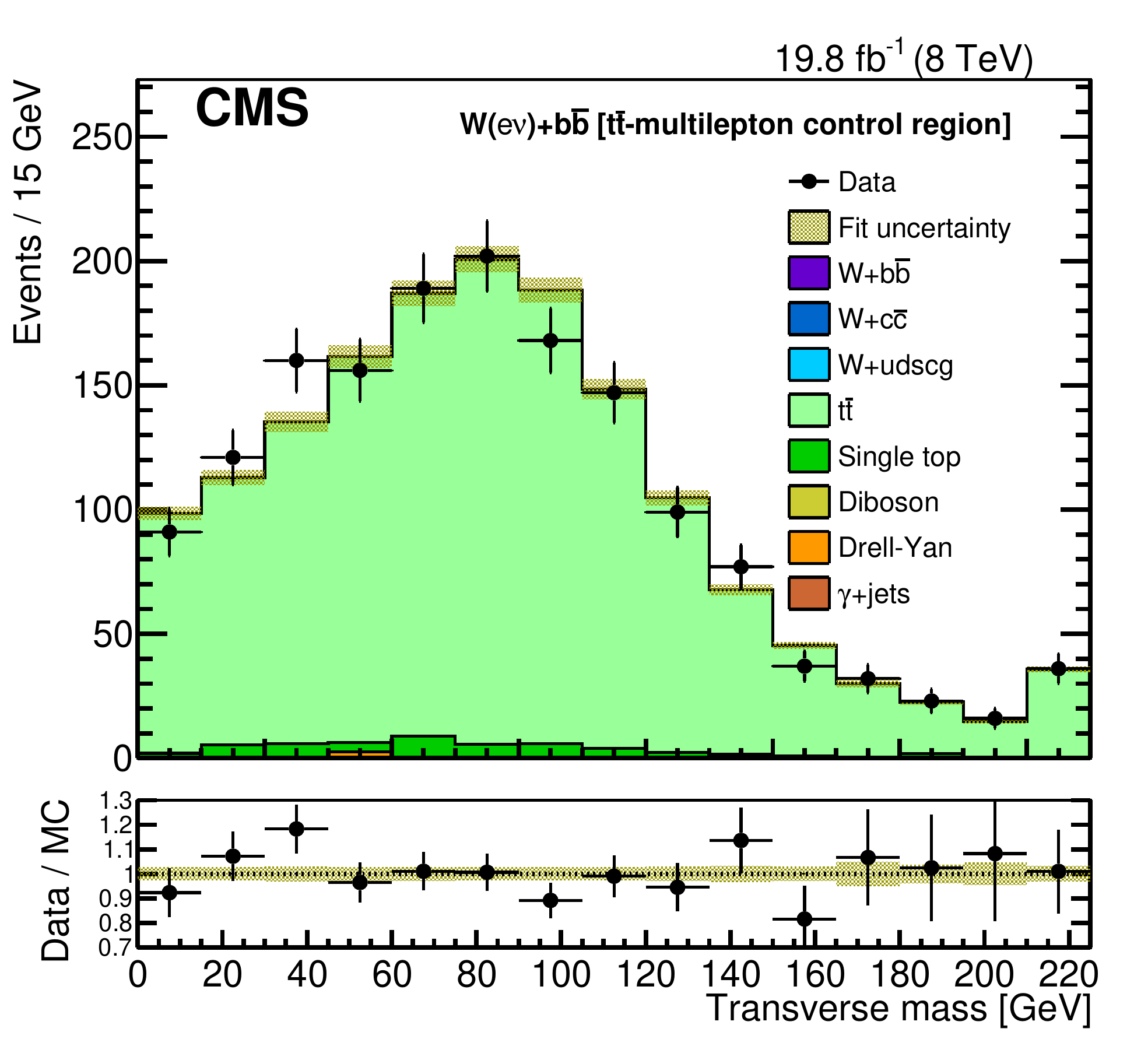}
\includegraphics[width=0.45\textwidth]{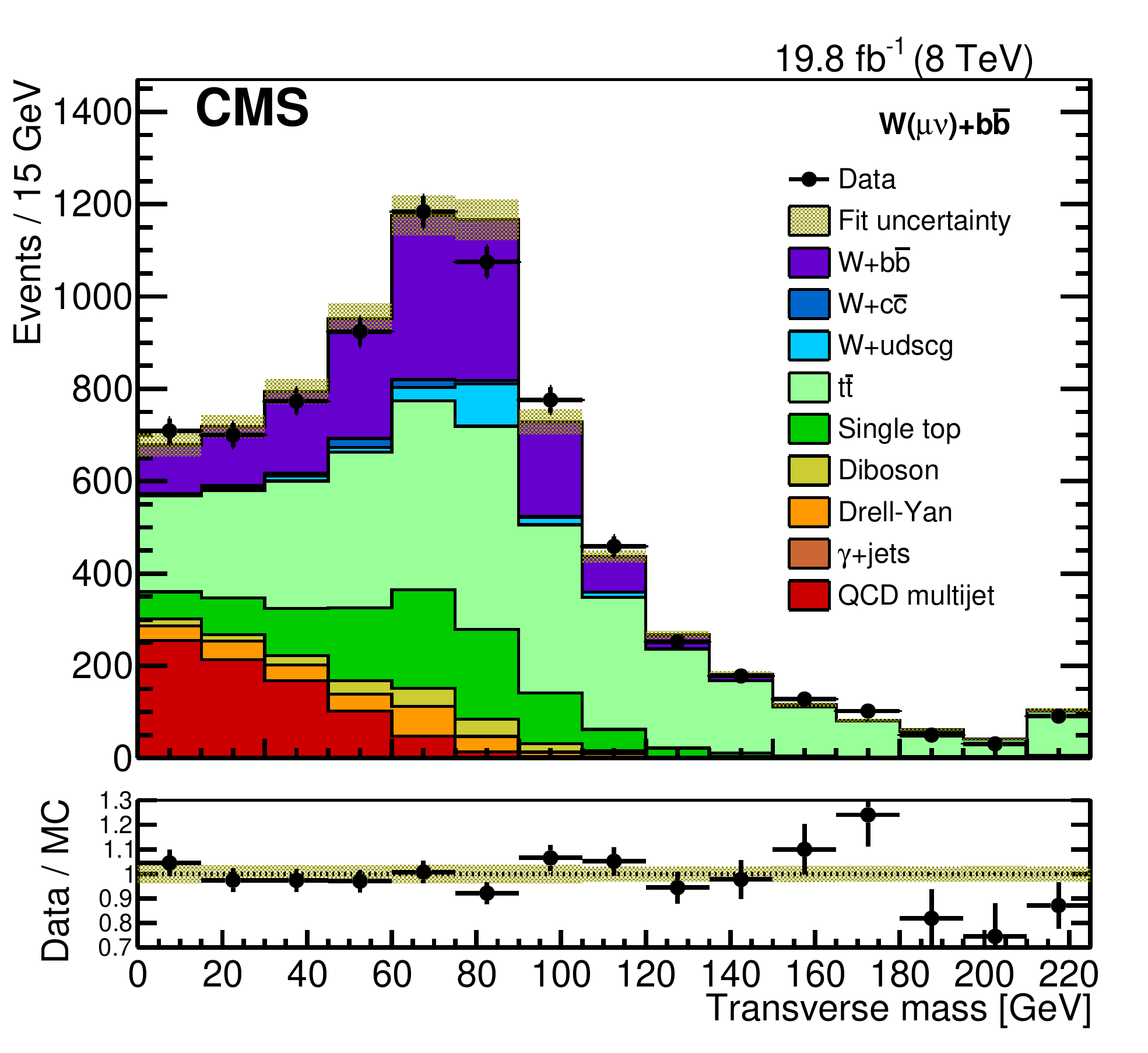}
\includegraphics[width=0.45\textwidth]{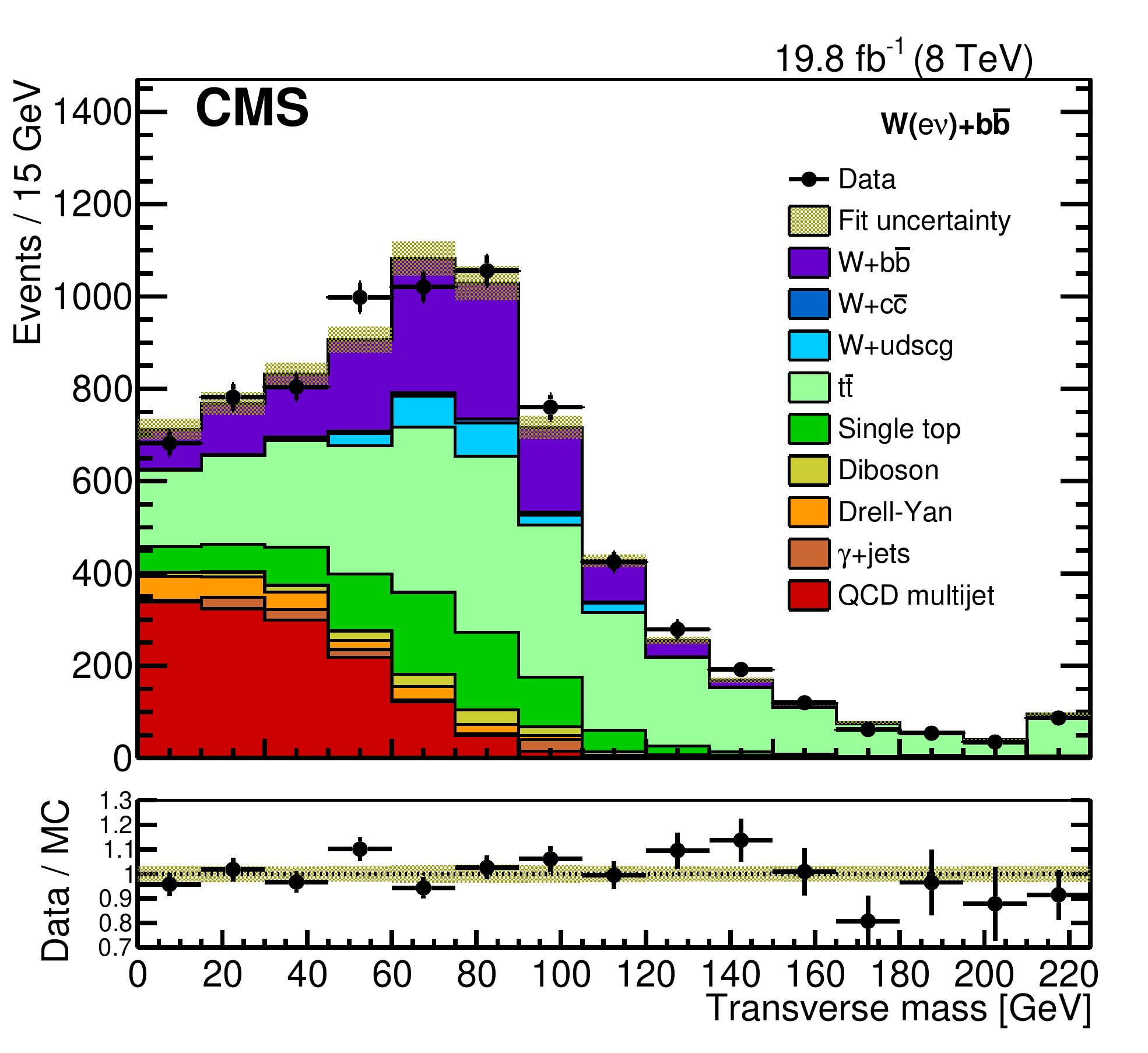}
\caption{
 The transverse mass distributions (upper) in the \ttbar-multijet
  phase space after fitting to obtain the b tagging efficiency rescale factors,
  (middle) in the \ttbar-multilepton sample
  after fitting to find the appropriate JES,
  and (lower) in the \Wbb signal sample
  after fitting simultaneously muon and electron decay channels.
 The lepton channels are shown separately with the muon
  sample on the left and the electron sample on the right.
 The last bin contains overflow events.
 The shaded area represents the total uncertainty in the simulation after the fit.
 }
\label{fig:fittingplots}
\end{figure*}

The fit in the $\ttbar$-multijet sample
 is used to obtain b tagging efficiency rescaling factors separately for
 the muon and electron channels
 in order to better describe the
 b tagging efficiency in the simulation
 as described in Section \ref{sec:analysis_strategy}.
The results of the fit are presented in the
 two plots at the top of Fig.~\ref{fig:fittingplots}.
The central values of the b tagging efficiency rescaling factors,
 {$1.12 \pm 0.08$} (muon channel) and
 {$1.16 \pm 0.08$} (electron channel), are averaged to {$1.14 \pm 0.08$}
 with the combined uncertainty, dominated by systematics,
 taken as the maximum of the
 uncertainties for the individual lepton channels.
The simulation is reweighted accordingly for the next fit, and
 the uncertainty in this fit sets the
 one standard deviation bound on the b tagging efficiency
 rescaling factor in subsequent fits.

A fit to the $\ttbar$-multilepton sample adjusts
 the JES, as described in Section \ref{sec:analysis_strategy}.
As a result, the simulated $\MT$ distributions
 change normalization.
The best fit results in changing the
 normalization by approximately 3.4\% from its central value,
 which corresponds to 1.3 standard deviations in JES.
The middle plots in Fig.~\ref{fig:fittingplots}
 show the results of the fits in the $\ttbar$-multilepton
 sample for
 the muon (left)
 and the electron (right) channels.
The JES is therefore shifted by 1.3 standard deviations in
 the simulation with the uncertainty
 taken from the fit.
Thus the simulation is tuned to describe the $\ttbar$
 control samples and is
 used to extract the signal yield in the signal region.

The results of the fit in the \Wbb signal region
 are shown in the bottom of Fig.~\ref{fig:fittingplots}.
All background contributions are allowed to vary
 in the fit within their uncertainties,
 while the \Wbb normalization remains a free parameter of the fit.
The correlations across all simulated samples are taken into account
 as shown in Table~\ref{tab:input_unc_wbb}.
Based on the fits the number of events of each
 type in the signal event sample is given in Table~\ref{tab:wbb_yields}.
Events coming from the production of a Higgs boson in association
 with a vector boson constitute a negligible fraction of the
 overall event yield and are not considered.

\begin{figure}[!htb]
\centering
\includegraphics[width=0.49\textwidth]{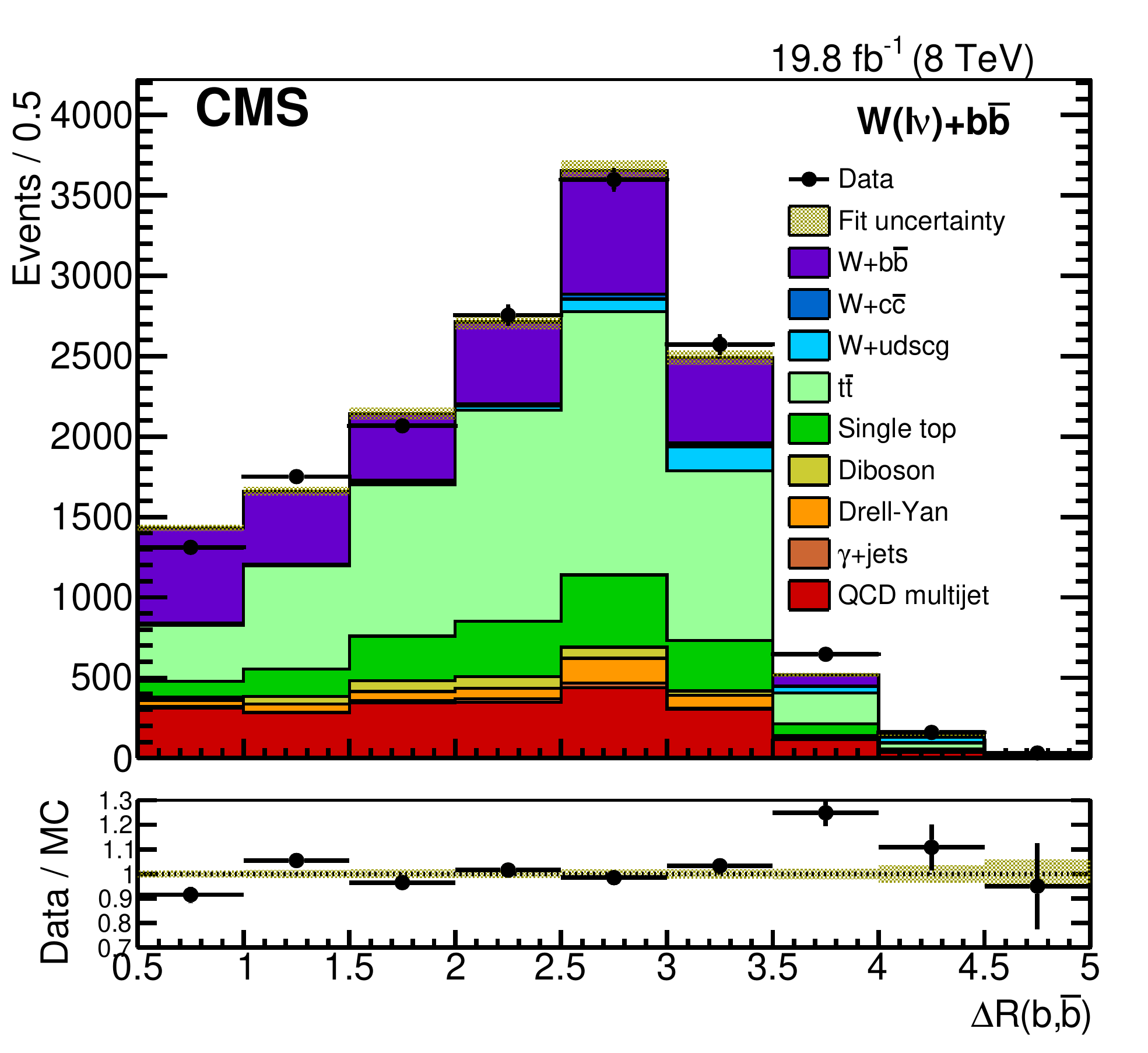}
\includegraphics[width=0.49\textwidth]{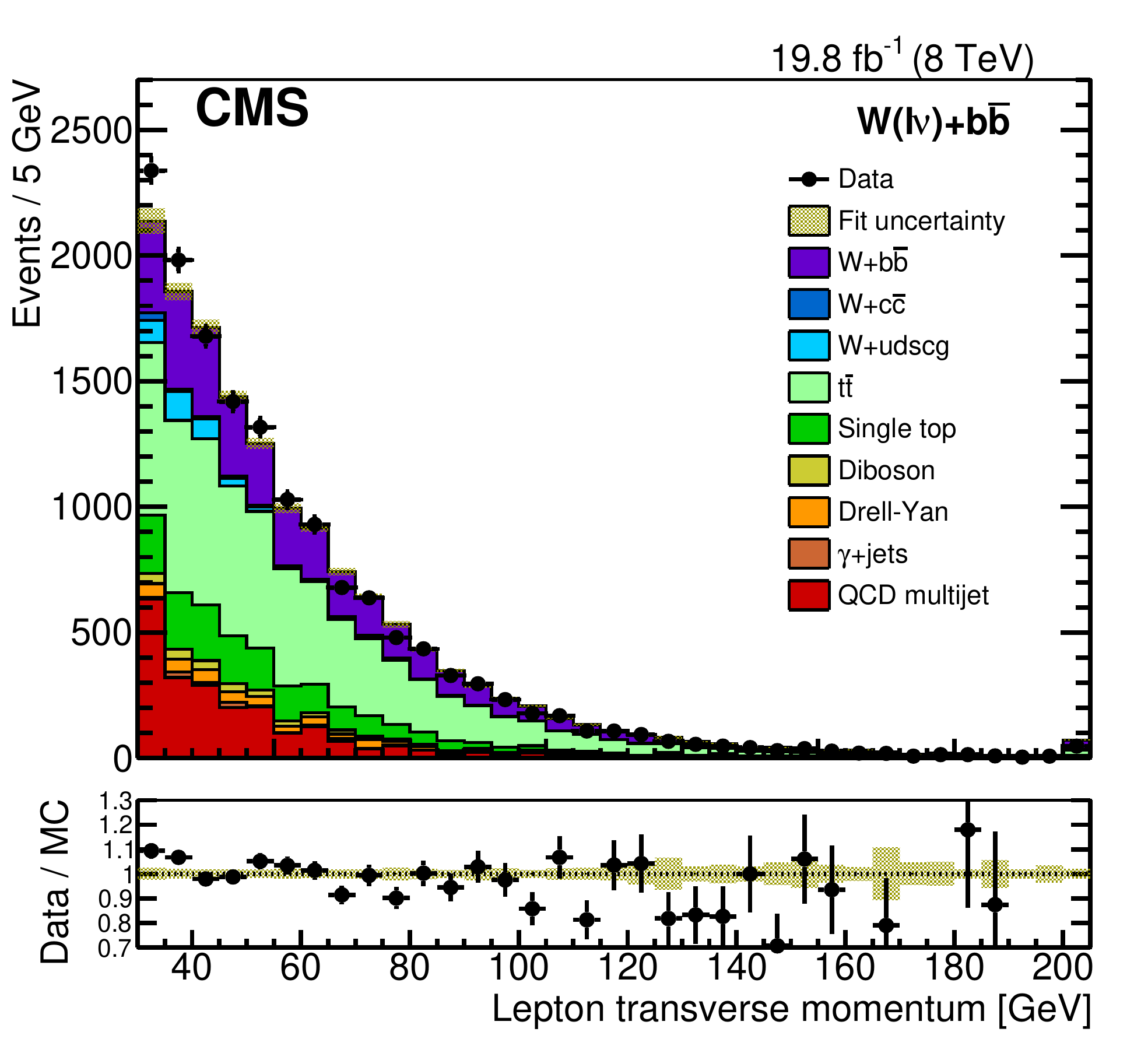}
\caption{
  Distributions of $\Delta R(\PQb,\PAQb)$ and $\pt^\ell$ after
   applying the results from the fits to the simulation.
  The QCD background shape is taken from an $\MT<30\GeV$ sideband and the
   muon and electron channels have been combined in these distributions.
  The last bin contains overflow events and
   the shaded area represents the total uncertainty in the simulation after the fit.
 }
\label{fig:postfit_drbb_ptl}
\end{figure}

Distributions for variables other than those being directly fit are
 also produced by applying the results from the three fits to
 the simulated samples.
Distributions of $\Delta R(\PQb,\PAQb)$ and $\pt^\ell$
 combining both lepton flavors are presented in Fig.~\ref{fig:postfit_drbb_ptl}.
The angular separation between the b jets
 is seen to be well modeled,
 and the $\PT^\ell$ distribution
 shows an agreement within $10\%$ for $\PT^\ell < 100$ GeV,
 with a slightly falling trend
 in the ratio of data and simulation.

\begin{table}[!htb]
\centering
\topcaption{
 Initial and final yields obtained in the \Wbb signal region.
 The uncertainties in the signal strength represent the
  total uncertainty of the fit.
}
\label{tab:wbb_yields}
 \begin{tabular}{r|r|r|r|r}
\multicolumn{1}{c}{}       & \multicolumn{2}{c|}{Muon}   & \multicolumn{2}{c}{Electron}   \\
\multicolumn{1}{c}{}       & \multicolumn{1}{c|}{Initial}      & Fitted      & Initial       & Fitted       \\
\hline \hline
Data     & \multicolumn{2}{c|}{7432}   & \multicolumn{2}{c}{7357}     \\
\hline & & & &
\\[-1em]
\Wbb          & 1323 & 1712 & 1121 &  1456 \\
\Wcc          &   60 &   61 &   36 &    37 \\
\Wudscg       &  182 &  179 &  220 &   217 \\
\ttbar        & 3049 & 3296 & 2640 &  2864 \\
Single top    &  958 & 1008 &  820 &   865 \\
Drell--Yan    &  261 &  265 &  220 &   224 \\
Diboson       &  175 &  181 &  139 &   144 \\
$\gamma+$jets &  --- &  --- &   98 &   105 \\
QCD           & 1109 &  803 & 1654 &  1373 \\
Total MC      & 7116 & 7505 & 6948 &  7284 \\
\hline
\hline
Signal strength & \multicolumn{2}{c|}{$1.21 \pm  0.19$} &  \multicolumn{2}{c}{$1.37 \pm 0.23$} \\
\hline
Combined & \multicolumn{4}{c}{$1.26 \pm 0.17$}  \\
 \end{tabular}

\end{table}

The cross section is calculated as
\ifthenelse{\boolean{cms@external}}{
\begin{multline*}
\sigma(\Pp\Pp\to \PW(\ell\nu)+\bbbar) =
\frac{N^{\text{data}}_{\text{reconstructed}}}{A\,\epsilon\, \mathcal{L}} \\=
\frac{N^{\text{data}}_{\text{reconstructed}}}{(N^{\mathrm{MC}}_{\text{reconstructed}}/N^{\mathrm{MC}}_{\text{generated}})\, \mathcal{L}} =
\alpha \sigma_{\text{gen}}
\end{multline*}
}{
\begin{equation*}
\sigma(\Pp\Pp\to \PW(\ell\nu)+\bbbar) =
\frac{N^{\text{data}}_{\text{reconstructed}}}{A\,\epsilon\, \mathcal{L}} =
\frac{N^{\text{data}}_{\text{reconstructed}}}{(N^{\mathrm{MC}}_{\text{reconstructed}}/N^{\mathrm{MC}}_{\text{generated}})\, \mathcal{L}} =
\alpha \sigma_{\text{gen}}
\end{equation*}
}
where
 $\mathcal{L}$ is the integrated luminosity,
 $N^{\text{data}}_{\text{reconstructed}}$ is the number of observed signal events,
 $N^{\mathrm{MC}}_{\text{reconstructed}}$ is the number of expected signal events from simulation
  reconstructed in the fiducial region,
 $N^{\mathrm{MC}}_{\text{generated}}$ is the number of generated events in the fiducial region,
 $A$ and $\epsilon$ are the acceptance and efficiency,
 $\alpha$ is the measured signal strength in the given lepton channel, and
 $\sigma_{\text{gen}}$ is the simulated fiducial cross section of the signal sample.
The signal strength is the scale factor
 in the \Wbb cross section predicted
 by the fit, after factoring out contributions to the
 overall change in normalization due to
 systematic effects which are correlated across samples.
In this analysis, the fiducial cross section is calculated as follows:
 {\MADGRAPH} is used to compute the \Wbb cross section with fiducial selections applied.
Then a k-factor for inclusive W production is applied
 that is obtained from the ratio
 of the inclusive W cross section calculated with
 \FEWZ~3.1 (at NNLO using the five-flavour
 CTEQ6M PDF set) and to that with \MADGRAPH.
The product $A\,\epsilon$ is 10 to 15\%
 and results from the
 combined effects of the efficiency of the
 lepton identification requirements (80\%) and b tagging efficiency (40\% per jet)
 and has an uncertainty of 10\%, arising from scale and PDF choices
 as indicated in the bottom of
 Table~\ref{tab:input_unc_wbb}.

The \Wbb cross section is measured within a fiducial volume, which is defined by
 requiring leptons with {$\pt > 30\GeV$}
 and {$\abs{\eta}<2.1$} and exactly two b-tagged jets of
 {$\pt > 25\GeV$} and {$\abs{\eta}<2.4$}.
The measured cross sections are presented in Table~\ref{tab:crosssections}.
The combination of the muon and electron measurements
 is done using a simultaneous fit to both channels,
 taking into account correlations across samples.

\begin{table*}[htbp]
\centering
\topcaption{
 Measured cross sections in the muon,
  electron, and combined lepton channels.
 The systematic uncertainty (syst)
 includes the contributions from all rows
 in Table~\ref{tab:input_unc_wbb} that
 have an entry in the ``Variation'' column,
 and the theoretical uncertainty (theo)
 includes the combination of the uncertainties
 associated with the choice of $\mu_{\mathrm{R}}$, $\mu_{\mathrm{F}}$, and PDF.
}
\label{tab:crosssections}
 \begin{tabular}{l|c}
 Channel        & $\sigma(\Pp\Pp\to \PW(\ell\nu)+\bbbar)\unit{pb}$ \\
\hline
 Combined & $  0.64 \pm 0.03\stat\pm 0.10\syst\pm 0.06\thy\pm 0.02\lum$ \\
\hline
     Muon & $  0.62 \pm 0.04\stat\pm 0.11\syst\pm 0.06\thy\pm 0.02\lum$ \\
 Electron & $  0.70 \pm 0.05\stat\pm 0.15\syst\pm 0.07\thy\pm 0.02\lum$ \\
 \end{tabular}
\end{table*}

The measured cross sections are compared to theoretical predictions from
 {\MCFM~7.0}~\cite{Campbell:2010ff, Badger:2010mg}
 with the {MSTW2008} PDF set, as well as from
 {\MADGRAPH~5}
 interfaced with \PYTHIA~6 in the four- and five-flavour schemes and
 {\MADGRAPH~5} with {\PYTHIA~8}~\cite{ref:Pythia8} in the four-flavour scheme.
In the four- and five-flavour approaches,
 the four and five lightest quark flavours are used
 in the proton
 PDF sets.
In the five-flavour scheme, the PDF set {CTEQ6L} is
 used and interfaced with {\PYTHIA6} using the {Z2*} tune.
The two four-flavour samples are produced using
 an NNLO PDF set interfaced with
 \PYTHIA version 6 using the {CTEQ6L} tune in one sample,
 and version 8 using the {CUETP8M1}
 tune~\cite{Khachatryan:2015pea} in the other.

Comparisons between the results of calculations performed
 under different assumptions provide important feedback
 on the validity of the techniques employed.
Differences in predictions arising from the modelling of
 b quarks as massive or massless are possible, as are
 variations in predictions arising from the use of different
 showering packages ({\PYTHIA6} vs.\ {\PYTHIA8})
 or matrix element
 generators ({\MADGRAPH} vs.\ {\MCFM~7.0}).
In the phase space explored here, these predictions are all
 very close in their central value and agree with each other
 well within their respective uncertainties.

The {\MCFM~7.0} cross section calculation is
 performed at the level of parton jets and thus
 requires a hadronization correction.
The multiplicative hadronization correction factor {$0.81\pm0.07$}
 is calculated using the {\MADGRAPH} + {\PYTHIA~6} sample
 and agrees well with
 the factor {$0.84\pm0.03$} calculated in
 the 7\TeV Z+b analysis~\cite{Chatrchyan:2014dha}.
The correction factor is obtained for
 jets computed excluding neutrinos from the particle list
 because such jets are closer in kinematics
 to particle jets at the detector level.
The uncertainty reflects both the limited statistics of the
 {\MADGRAPH} + {\PYTHIA~6} sample as well as a comparison with the
 {\MADGRAPH} + {\PYTHIA~8} sample.

The \MCFM~7.0 and four-flavour \MADGRAPH predictions do not
 take into account
 \Wbb production where the \bbbar system is
 produced in a different partonic level interaction
 than the one which produced the \PW~boson, albeit in the
 same collision.
Simulations of \MADGRAPH + \PYTHIA events that include
 double parton interactions (DPI) reproduce the \Wjets data~\cite{Chatrchyan:2013xxa}.
Therefore a \MADGRAPH + {\PYTHIA~8} sample of a \PW~boson produced in association with a
 \bbbar pair coming from DPI
 is generated to study the effect on the fiducial cross section.
Using this dedicated sample, an additive correction $\sigma_{\mathrm{DPI}}$
 is estimated to be {$0.06\pm0.06\unit{pb}$}, where the uncertainty
 is conservatively assigned to be 100\% of the value.

\begin{figure}[htbp]
\centering
\includegraphics[width=\cmsFigWidth]{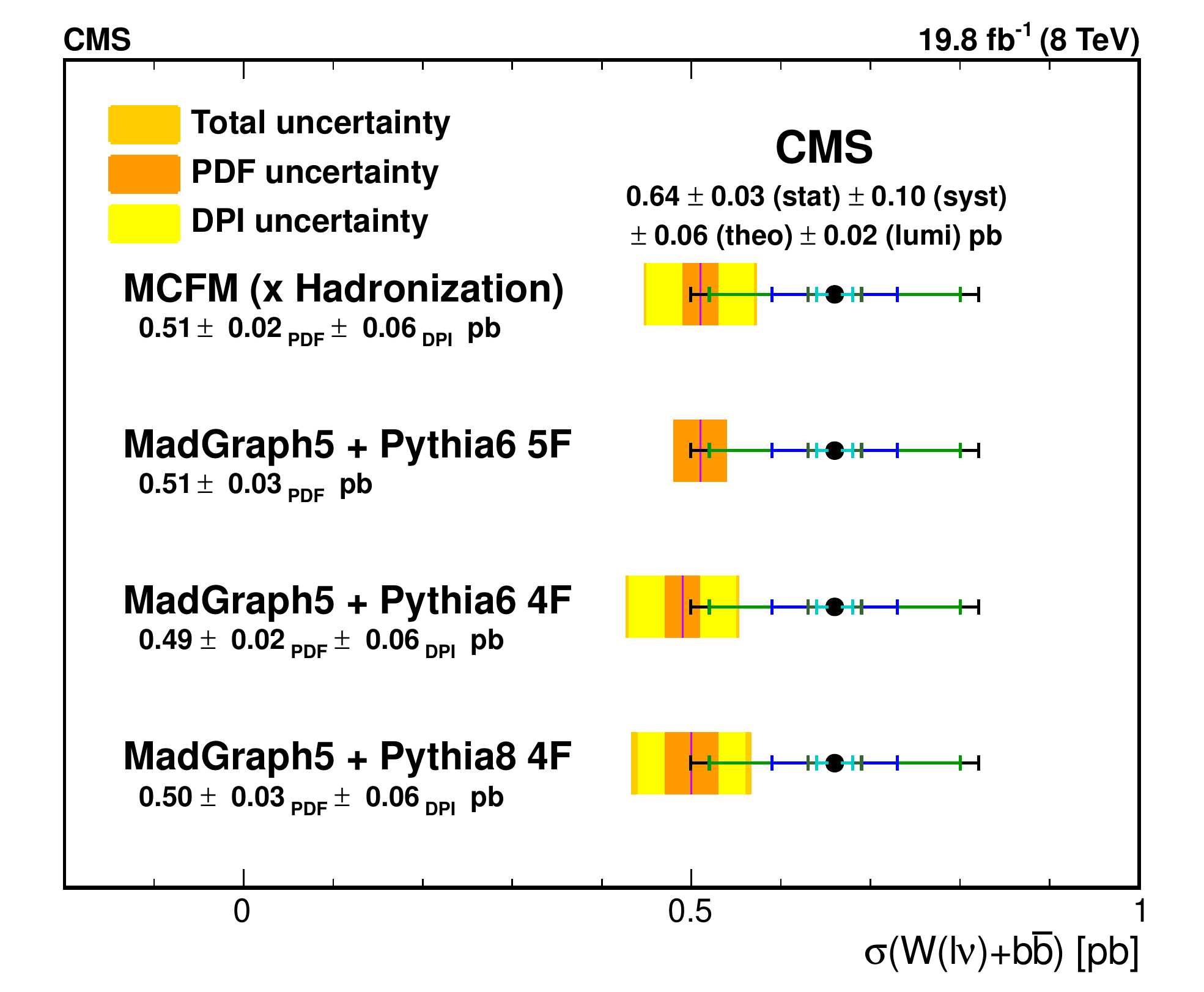}
\caption{
 Comparison between the measured \Wlnbb cross section and
  various QCD predictions.
 The orange band indicates the uncertainty
  in the given sample associated with PDF choice
  and the yellow band represents the uncertainty
  associated with DPI.
 The labels 4F and 5F refer to the four- and five-flavour
  PDF schemes.
 In the case of the \MADGRAPH + {\PYTHIA6} (5F) sample,
  the effects of DPI are already included in the generated
  samples so the DPI correction is not needed.
 The measured cross section is also shown with
  the total uncertainty in
  black and the luminosity, statistical, theoretical, and systematic
  uncertainties indicated.
 }
\label{fig:xc_comparison}
\end{figure}

The resulting cross section predictions in the fiducial
 phase space at the hadron level, including the estimated
 hadronization and DPI corrections as needed,
 are compared in Fig.~\ref{fig:xc_comparison}
 with the measured value.
Within one standard deviation the predictions agree with the measured cross section.

\clearpage

\section{Summary}
The cross section for the production of a W~boson
 in association
 with two b jets was measured using a
 sample of proton-proton collisions at
 $\sqrt{s} = 8\TeV$ collected by the CMS experiment.
The data sample corresponds to an integrated
 luminosity of 19.8\fbinv.
The W~bosons were reconstructed via their
 leptonic decays, {$\PW\to\ell\nu$},
 where  {$\ell=\mu$ or \Pe}.
The fiducial region studied contains exactly one
 lepton with transverse momentum $\pt^{\ell}>30\GeV$
 and pseudorapidity {$\abs{\eta^{\ell}}<2.1$},
 with exactly two b jets with $\pt>25\GeV$
 and {$\abs{\eta}<2.4$}
 and no other jets with $\pt>25\GeV$
 and {$\abs{\eta}<4.7$}.
The cross section is
 {$\sigma ( {{\Pp\Pp}} \to {\PW} (\ell\nu)+\bbbar)= 0.64 \pm 0.03\stat\pm 0.10\syst\pm 0.06\thy\pm 0.02\lum\unit{pb}$},
 in agreement with standard model predictions.

\begin{acknowledgments}
\hyphenation{Rachada-pisek}
We congratulate our colleagues in the CERN accelerator departments for the excellent performance of the LHC and thank the technical and administrative staffs at CERN and at other CMS institutes for their contributions to the success of the CMS effort. In addition, we gratefully acknowledge the computing centers and personnel of the Worldwide LHC Computing Grid for delivering so effectively the computing infrastructure essential to our analyses. Finally, we acknowledge the enduring support for the construction and operation of the LHC and the CMS detector provided by the following funding agencies: BMWFW and FWF (Austria); FNRS and FWO (Belgium); CNPq, CAPES, FAPERJ, and FAPESP (Brazil); MES (Bulgaria); CERN; CAS, MoST, and NSFC (China); COLCIENCIAS (Colombia); MSES and CSF (Croatia); RPF (Cyprus); SENESCYT (Ecuador); MoER, ERC IUT and ERDF (Estonia); Academy of Finland, MEC, and HIP (Finland); CEA and CNRS/IN2P3 (France); BMBF, DFG, and HGF (Germany); GSRT (Greece); OTKA and NIH (Hungary); DAE and DST (India); IPM (Iran); SFI (Ireland); INFN (Italy); MSIP and NRF (Republic of Korea); LAS (Lithuania); MOE and UM (Malaysia); BUAP, CINVESTAV, CONACYT, LNS, SEP, and UASLP-FAI (Mexico); MBIE (New Zealand); PAEC (Pakistan); MSHE and NSC (Poland); FCT (Portugal); JINR (Dubna); MON, RosAtom, RAS and RFBR (Russia); MESTD (Serbia); SEIDI and CPAN (Spain); Swiss Funding Agencies (Switzerland); MST (Taipei); ThEPCenter, IPST, STAR and NSTDA (Thailand); TUBITAK and TAEK (Turkey); NASU and SFFR (Ukraine); STFC (United Kingdom); DOE and NSF (USA).

Individuals have received support from the Marie-Curie program and the European Research Council and EPLANET (European Union); the Leventis Foundation; the A. P. Sloan Foundation; the Alexander von Humboldt Foundation; the Belgian Federal Science Policy Office; the Fonds pour la Formation \`a la Recherche dans l'Industrie et dans l'Agriculture (FRIA-Belgium); the Agentschap voor Innovatie door Wetenschap en Technologie (IWT-Belgium); the Ministry of Education, Youth and Sports (MEYS) of the Czech Republic; the Council of Science and Industrial Research, India; the HOMING PLUS program of the Foundation for Polish Science, cofinanced from European Union, Regional Development Fund, the Mobility Plus program of the Ministry of Science and Higher Education, the National Science Center (Poland), contracts Harmonia 2014/14/M/ST2/00428, Opus 2013/11/B/ST2/04202, 2014/13/B/ST2/02543 and 2014/15/B/ST2/03998, Sonata-bis 2012/07/E/ST2/01406; the Thalis and Aristeia programs cofinanced by EU-ESF and the Greek NSRF; the National Priorities Research Program by Qatar National Research Fund; the Programa Clar\'in-COFUND del Principado de Asturias; the Rachadapisek Sompot Fund for Postdoctoral Fellowship, Chulalongkorn University and the Chulalongkorn Academic into Its 2nd Century Project Advancement Project (Thailand); and the Welch Foundation, contract C-1845.
\end{acknowledgments}

\bibliography{auto_generated}

\cleardoublepage \appendix\section{The CMS Collaboration \label{app:collab}}\begin{sloppypar}\hyphenpenalty=5000\widowpenalty=500\clubpenalty=5000\textbf{Yerevan Physics Institute,  Yerevan,  Armenia}\\*[0pt]
V.~Khachatryan, A.M.~Sirunyan, A.~Tumasyan
\vskip\cmsinstskip
\textbf{Institut f\"{u}r Hochenergiephysik der OeAW,  Wien,  Austria}\\*[0pt]
W.~Adam, E.~Asilar, T.~Bergauer, J.~Brandstetter, E.~Brondolin, M.~Dragicevic, J.~Er\"{o}, M.~Flechl, M.~Friedl, R.~Fr\"{u}hwirth\cmsAuthorMark{1}, V.M.~Ghete, C.~Hartl, N.~H\"{o}rmann, J.~Hrubec, M.~Jeitler\cmsAuthorMark{1}, A.~K\"{o}nig, I.~Kr\"{a}tschmer, D.~Liko, T.~Matsushita, I.~Mikulec, D.~Rabady, N.~Rad, B.~Rahbaran, H.~Rohringer, J.~Schieck\cmsAuthorMark{1}, J.~Strauss, W.~Treberer-Treberspurg, W.~Waltenberger, C.-E.~Wulz\cmsAuthorMark{1}
% ...
\vskip\cmsinstskip
% ...
1:~~Also at Vienna University of Technology, Vienna, Austria\\
% ...

\end{sloppypar}
\end{document}